# Relative Entropy Regularised TDLAS Tomography for Robust Temperature Imaging

Yong Bao[1], *Member, IEEE*, Rui Zhang[1], Godwin Enemali[1], Zhang Cao[2], *IEEE Member*, Bin Zhou[3],
Hugh McCann[1] and Chang Liu[1]*, *Member, IEEE*
1 School of Engineering, University of Edinburgh, Edinburgh EH9 3JL, U.K.
2 School of Instrumentation and Optoelectronic Engineering, Beihang University, Beijing, China
3 School of Energy and Environment, Southeast University, Nanjing, China
*Corresponding author, E-mail: C.Liu@ed.ac.uk

*Abstract* — Tunable Diode Laser Absorption Spectroscopy (TDLAS) tomography has been widely used for *in situ* combustion diagnostics, yielding images of both species concentration and temperature. The temperature image is generally obtained from the reconstructed absorbance distributions for two spectral transitions, i.e. two-line thermometry. However, the inherently ill-posed nature of tomographic data inversion leads to noise in each of the reconstructed absorbance distributions. These noise effects propagate into the absorbance ratio and generate artefacts in the retrieved temperature image. To address this problem, we have developed a novel algorithm, which we call Relative Entropy Tomographic RecOnstruction (RETRO), for TDLAS tomography. A relative entropy regularisation is introduced for high-fidelity temperature image retrieval from jointly reconstructed two-line absorbance distributions. We have carried out numerical simulations and proof-of-concept experiments to validate the proposed algorithm. Compared with the well-established Simultaneous Algebraic Reconstruction Technique (SART), the RETRO algorithm significantly improves the quality of the tomographic temperature images, exhibiting excellent robustness against TDLAS tomographic measurement noise. RETRO offers great potential for industrial field applications of TDLAS tomography, where it is common for measurements to be performed in very harsh environments.

Keywords: laser absorption spectroscopy; tomography; relative entropy; regularisation; two-line thermometry.

## I. INTRODUCTION

Since the first experimental demonstration of high-speed chemical species tomography [1], interest in Tunable Diode Laser Absorption Spectroscopy (TDLAS) tomography has grown rapidly, with a focus on non-invasive imaging of critical combustion parameters, e.g. temperature [2-5], gas concentration [3-7], pressure [8, 9] and velocity [10], in reactive flows. High-fidelity temperature imaging is of critical interest, as the temperature distribution directly relates to heat transfer and reveals combustion efficiency and temperature-dependent creation of pollutants, such as NOx and CO. Therefore, many papers on TDLAS tomography have focused on improving the accuracy and robustness of temperature imaging.

The widely adopted method in TDLAS tomography for temperature imaging is the so-called two-line strategy [11], where the absorbance distributions for two spectral transitions with different temperature-dependent line strengths are individually reconstructed, then the temperature image is retrieved from the ratio of the absorbances in each pixel of the Region of Interest (RoI). Alternatively, temperature images can be reconstructed by using spectra for multiple transitions, the so-called hyperspectral tomography (HT) [12, 13]. Although better accuracy and noise resistance can be achieved by HT, measurement of the necessary spectra requires expensive hardware, e.g. broadband lasers and detectors [5], and introduces high computational cost for data inversion. In contrast, the two-line strategy is cost-effective for temperature retrieval, placing minimal requirements on optical components and computing resources, and it is adopted in this paper.

Temperature imaging using the two-line strategy requires solution of two linear inverse problems by reconstructing, from the line-of-sight (LOS) TDLAS measurements, the absorbance distribution for each spectral feature. The inverse problems are inherently ill-posed due to the limited number of available LOS measurements and their inevitable uncertainties, leading to noise in each of the reconstructed absorbance distributions [14]. To mitigate the instability of the reconstructions, a variety of computational tomographic algorithms have been applied in TDLAS tomography, e.g. Algebraic Reconstruction Technique (ART) [15], Simultaneous Algebraic Reconstruction Technique (SART) [3, 16], Landweber algorithm [4, 17], and Tikhonov regularisation [18], by formulating the inverse problem with a heuristically determined prior, e.g. smoothness of the true temperature distributions. Noise in the retrieved absorbance distributions, particularly in the denominator of the two-line ratio, propagates into the ratio and generates spike noise in the temperature image. To address this problem, we have developed a tomographic reconstruction algorithm using relative entropy regularisation, and the algorithm is called Relative Entropy Tomographic RecOnstruction (RETRO). The inverse problem of RETRO is solved by cone optimisation, enabling high-fidelity temperature retrieval from jointly reconstructed two-line absorbance distributions. As well as using smoothness constraints, the relative entropy regularisation further alleviates the spike noise problem arising from the individually reconstructed two-line absorbance distributions. The proposed RETRO algorithm promises to be highly beneficial for industrial applications of TDLAS tomography, e.g. gas turbine exhaust imaging [19] and power-plant boiler diagnosis [20], where significant noise contamination of TDLAS measurements is common due to the harsh measurement environment. Recently, statistical inversion has been investigated by imposing *a priori* knowledge on the desired solution based on the framework of Bayesian statistics. The solution is then calculated based on the maximum *a posteriori* (MAP) estimation [21] or covariance estimation [22]. This paper does not address statistical inversion techniques, due in part to their high computational cost and their susceptibility to inadequate prior models [16].

## II. METHODS

### A. Formulation of TDLAS tomography

When a laser beam at frequency $v$ [cm$^{-1}$] penetrates an absorbing gas sample on a path of length $L$ [cm], a proportion of its intensity is absorbed. According to Beer's law, the wavelength-dependent absorbance can be defined as

$$\alpha(v) = \ln \frac{I_0(v)}{I_t(v)} = \int_0^L P(l)X(l)S(T(l))\phi(v,l)dl, \quad (1)$$

where $I_0(v)$ and $I_t(v)$ are the incident and transmitted laser intensities, respectively, and $l$ is the position along the path. $P(l)$ [atm] is the local total pressure, $X(l)$ the local molar fraction of the absorbing species, $T(l)$ [K] the local temperature, $\phi(v, l)$ [cm] the line-shape function, and $S(T)$ [cm$^{-2}$atm$^{-1}$] the temperature-dependent line strength.

Since the line-shape function can be normalized as $\int_{-\infty}^{+\infty} \phi(v,l)dv \equiv 1$, the path integral $A_v$ can be formulated as:

$$A_v = \int_{-\infty}^{\infty} \alpha(v)dv = \int_0^L PX(l)S_v(T(l))dl = \int_0^L a_v(l)dl, \quad (2)$$

where $a_v$ is the local density of the path integral.

For TDLAS tomography, Eqn. (2) can be discretised as:

$$A_v = La_v, \quad (3)$$

where $A_v \in \mathbb{R}^{M \times 1}$ denotes the vector of path integrals obtained from $M$ LOS-TDLAS measurements. $L \in \mathbb{R}^{M \times N}$ is the sensitivity matrix with its element $l_{i,j}$ representing the length of the laser path segment for the $i$-th laser beam passing through the $j$-th pixel. $i$ ($i=1, 2..., M$) and $j$ ($j=1, 2..., N$) are the indices of laser beams and pixels, respectively. $a_v \in \mathbb{R}^{N \times 1}$ is the vector of the density of $A_v$ with its elements $a_{v,j} = P_j X_j S(T_j)$.

### B. SART-based TDLAS Tomographic reconstruction

SART is one of the most well-established algorithms for hard-field tomography. This technique maintains the rapid convergence rate of the ART, while retaining the noise-suppression features of the Simultaneous Iterations Reconstruction Technique (SIRT) [23]. In this paper, SART is used as a representative tomographic algorithm and its performance is compared with the proposed RETRO.

Using the *a priori* information of smoothness, the two-line absorbance distributions, $a_{v1}$ and $a_{v2}$, can be reconstructed by solving the following regularised least squares problem

$$\arg\min_{a_v} \left\{ \|A_v - La_v\|_W^2 + \lambda \|Fa_v\|_2^2 \right\} \; s.t. \; a_v \geq 0, \quad (4)$$

where $\|Fa_v\|_2^2$ is the first-order Tikhonov regularisation term with a linear differential operator $F$. $\lambda$ is the empirically determined regularisation parameter. The residual term $\|A_v - La_v\|_W^2$ is weighted per unit length of the laser path using $W$, which is defined as

$$W = diag(1/l_{ray,1}, 1/l_{ray,2}, ..., 1/l_{ray,M}), \quad (5)$$

where $l_{ray,i}$ denotes the length of the $i$-th beam through the RoI

$$l_{ray,i} = \sum_{j=1}^{N} l_{i,j}. \quad (6)$$

By utilising adaptive step size $\eta$, and non-negative projection operator $\Pi_+()$, SART solves Eqn. (4) iteratively as

$$a_v^{k+1} = \Pi_+ \left( a_v^k + \eta C L^T W (A_v - La_v^k) - \lambda C F^T F a_v^k \right), \quad (7)$$

where $\Pi_+()$ operates as $\Pi_+(a_{v,j}) = \max(0, a_{v,j})$.

$C$ is the diagonal preconditioner defined as:

$$C = diag(1/l_{pixel,1}, 1/l_{pixel,2}, ..., 1/l_{pixel,N}), \quad (8)$$

where

$$l_{pixel,j} = \sum_{i=1}^{M} l_{i,j}. \quad (9)$$

In each iteration, $\eta$ is updated by back-tracking line search [24].

Finally, the temperature in the $j$-th pixel, $T_j$, can be calculated by the two-line absorbance ratio, $R_j$, given by

$$R_j = \frac{a_{v2,j}}{a_{v1,j}} = \frac{S_{v2}(T_j)}{S_{v1}(T_j)}. \quad (10)$$

Eqn. (10) shows that the quality of the temperature image directly depends on the noise level of the two-line absorbance ratio. Although smoothness regularisation is used in SART, the individually reconstructed $a_{v1}$ and $a_{v2}$ inevitably suffer from perturbations caused by the ill-posed nature of TDLAS tomography. Fig. 1 illustrates intuitively the resulting spike noise for each pixel in the temperature image. For these calculations, the inverse problem is formulated using the tomographic sensor and phantom 2 introduced in Section III A below; TDLAS measurements are simulated with a signal-to-noise ratio (SNR) of 40 dB. The reconstructed and original $a_{v1,j}$ and $a_{v2,j}$ with their relative errors $e_{v1,j}$ and $e_{v2,j}$ in each pixel are shown in Figs. 1 (a) and (b), respectively. Although the maximum values of $e_{v1}$ and $e_{v2}$ are 0.046 and 0.103 respectively, very large spike noise effects can be observed in the $T_j$ values shown in Fig. 1 (c), with a maximum relative error $e_{T,j}$ of 1.51.

### C. RETRO-based TDLAS tomographic reconstruction

The RETRO algorithm proposed here introduces a constraint that directly enforces smoothness of the two-line absorbance ratio. Both $a_{v1}$ and $a_{v2}$ are jointly reconstructed by solving the following inverse problem:

$$\min_{a_{v1}, a_{v2}} \left\{ \left\| \begin{bmatrix} A_{v1} \\ 0 \end{bmatrix} - \begin{bmatrix} La_{v1} \\ \gamma Fa_{v1} \end{bmatrix} \right\|_2^2 + \left\| \begin{bmatrix} A_{v2} \\ 0 \end{bmatrix} - \begin{bmatrix} La_{v2} \\ \gamma Fa_{v2} \end{bmatrix} \right\|_2^2 \right\} \quad (11)$$

$$s.t. \; a_{v1} > 0, \; a_{v2} > 0, \; \|a_{v2}./a_{v1}\|_2^2 < r,$$

where $\gamma$ is the smoothness regularisation parameter, and $r$ the ratio constraint parameter.

The constrained optimisation problem in Eqn. (11) can be reformulated into the following unconstrained problem using Lagrange's multipliers,

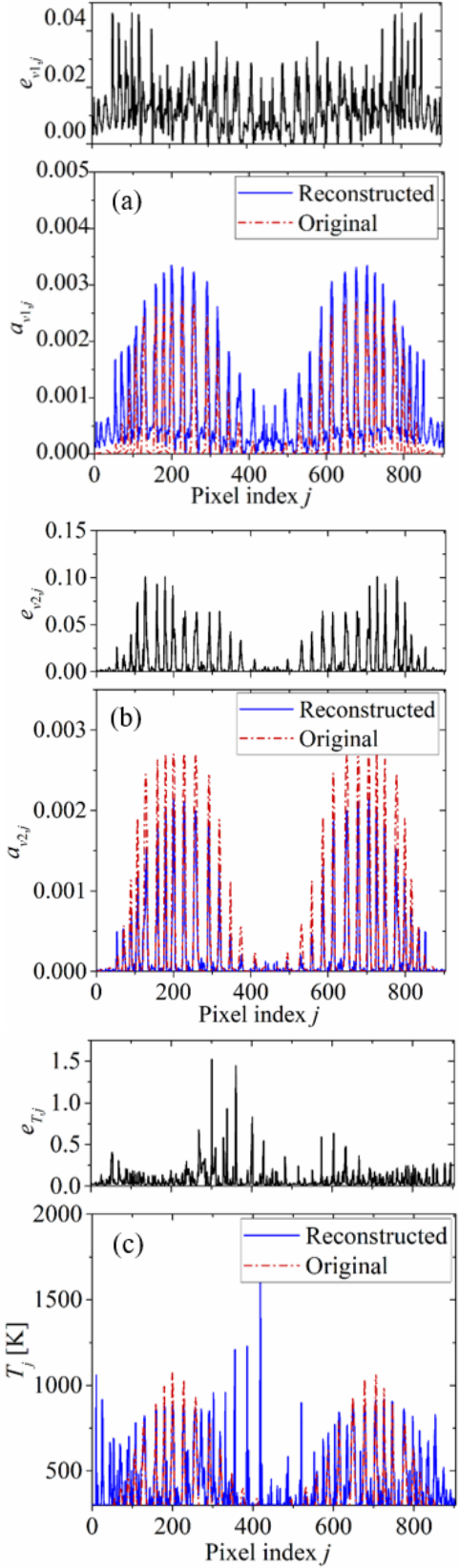

Figure 1: Spike noise introduced in temperature reconstruction using SART. (a), (b) and (c) show the reconstructed and original $a_{v1,j}$, $a_{v2,j}$, and $T_j$ with their relative errors $e_{v1,j}$; $e_{v2,j}$; and $e_{T,j}$, respectively.

$$\min_{a_{v1},a_{v2}} \left\{ \left\| \begin{bmatrix} A_{v1} \\ 0 \end{bmatrix} - \begin{bmatrix} La_{v1} \\ \gamma Fa_{v1} \end{bmatrix} \right\|_2^2 + \left\| \begin{bmatrix} A_{v2} \\ 0 \end{bmatrix} - \begin{bmatrix} La_{v2} \\ \gamma Fa_{v2} \end{bmatrix} \right\|_2^2 + \mu \| a_{v2}./a_{v1} \|_2^2 \right\}$$ (12)

$$s.t.\ a_{v1} > 0,\ a_{v2} > 0,$$

where $\mu$ represents the two-line absorbance ratio regularisation parameter.

However, the regularisation term $\mu \|a_{v2}./a_{v1}\|_2^2$ is not convex [25], therefore it is generally difficult to use in determination of the globally optimal solution for Eqn. (12) [26]. To replace this regularisation term, we developed a relative entropy function $g(a_{v2}, a_{v1})$ that is jointly convex in terms of both $a_{v2}$ and $a_{v1}$ [27], as follows:

$$g(a_{v2}, a_{v1}) = \begin{cases} (a_{v2} + a_{v1}) \circ \log(1 + a_{v2}./a_{v1}) & \text{if } a_{v2}, a_{v1} > 0 \\ 0 & \text{if } a_{v1} + a_{v2} = 0,\ a_{v1} \geq 0 \\ +\infty & \text{otherwise} \end{cases}$$ (13)

The proposed relative entropy regularisation term $\|(a_{v2} + a_{v1}) \circ \log(1 + a_{v2}./a_{v1})\|_2^2$ in Eqn. (13) has two main advantages: a) it is convex and therefore enables an unique solution; b) it balances well the penalty on a wide range of $a_{v2}./a_{v1}$. For any given $j$, the imposed penalty from the Logarithmic term $\log^2(1+a_{v2,j}/a_{v1,j})$ is calculated with different values of $a_{v2,j}/a_{v1,j}$. As shown in Fig. 2, $\log^2(1+a_{v2,j}/a_{v1,j})$ is close to zero given small values of $a_{v2,j}/a_{v1,j}$, denoting the regularisation has minor influence on typical $a_{v2,j}/a_{v1,j}$ that are generally small. As $a_{v2,j}/a_{v1,j}$ increases, the imposed regularisation becomes stronger in order to suppress large perturbations in the retrieved temperature, such as the spike noise shown in Fig. 1. Therefore, it can be seen that the relative entropy regularisation developed above is superior for keeping useful details of the reconstructed image while maintaining good performance on spike noise suppression.

By applying the relative entropy function $g(a_{v2}, a_{v1})$ in Eqn. (12), and retaining the empirical tuning parameter $\mu$, the inverse problem of TDLAS tomography can be formulated as

$$\min_{a_{v1},a_{v2}} \left\{ \left\| \begin{bmatrix} A_{v1} \\ 0 \end{bmatrix} - \begin{bmatrix} La_{v1} \\ \gamma Fa_{v1} \end{bmatrix} \right\|_2^2 + \left\| \begin{bmatrix} A_{v2} \\ 0 \end{bmatrix} - \begin{bmatrix} La_{v2} \\ \gamma Fa_{v2} \end{bmatrix} \right\|_2^2 + \mu \|(a_{v2} + a_{v1}) \circ \log(1 + a_{v2}./a_{v1})\|_2^2 \right\}$$ (14)

$$s.t.\ a_{v1} > 0,\ a_{v2} > 0.$$

In general, the inverse problem in Eqn. (14) can be solved by successive approximation of the convex regularisation term $\|(a_{v2} + a_{v1}) \circ \log(1 + a_{v2}./a_{v1})\|_2^2$, which suffers from very high computational cost [28]. Alternatively, we recast Eqn. (14) as a conic optimisation problem

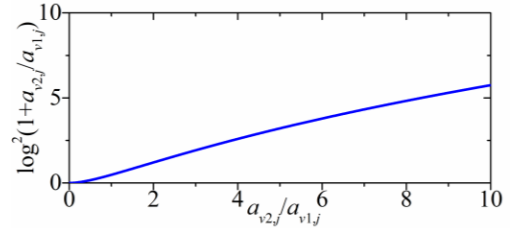

Figure 2: Dependence of $\log^2(1+ a_{v2,j}/a_{v1,j})$ on $a_{v2,j}/a_{v1,j}$.

$$\min_{a_{v1}, a_{v2}, \tau} \left\{ \left\| \begin{bmatrix} A_{v1} \\ 0 \end{bmatrix} - \begin{bmatrix} La_{v1} \\ \gamma Fa_{v1} \end{bmatrix} \right\|_2^2 + \left\| \begin{bmatrix} A_{v2} \\ 0 \end{bmatrix} - \begin{bmatrix} La_{v2} \\ \gamma Fa_{v2} \end{bmatrix} \right\|_2^2 + \mu \sum_{j=1}^{N} \tau_j \right\} \quad (15)$$

$$s.t.\ a_{v1} > 0,\ a_{v2} > 0,\ (a_{v1} + a_{v2}, a_{v1}, \tau) \in K_{re},$$

where $K_{re}$ is the (scalar) relative entropy cone defined by

$$K_{re} \equiv \left\{ (x, y, \tau) \in \mathbb{R}\ \ \mathbb{R}\ \ \mathbb{R}\ \ : (x_j / y_j) \leq \tau_j, \forall j \right\} \quad (16)$$

and $\tau \in \mathbb{R}_{++}^N$ the auxiliary vector.

Consequently, $a_{v1}$ and $a_{v2}$ in Eqn. (16) can be solved jointly and efficiently using the interior-point method [29], which is implemented by the MOSEK optimization suite for MATLAB [30]. With $a_{v1}$ and $a_{v2}$ in hand, the temperature image can be reconstructed by Eqn. (10).

### III. NUMERICAL SIMULATION

In this section, the proposed RETRO algorithm is validated using numerical simulations of three phantoms with different numbers of inhomogeneities. The quality of the tomographic image was quantitatively examined in terms of image error, dislocation, accuracy of centroid value and overshoot. To demonstrate the superiority of the proposed algorithm, its performance was compared with SART.

#### A. Simulation setup

Water vapour ($H_2O$) has a strong near-infrared absorption spectrum and is a principal product of hydrocarbon combustion. Therefore, it is selected as the target absorption species to demonstrate RETRO for TDLAS tomographic temperature imaging. The absorption transitions at $v_1 = 7185.6$ cm$^{-1}$ and $v_2 = 7444.36$ cm$^{-1}$ are used in this work as they have moderate line strengths and good temperature sensitivity over 300 - 1200 K [31]. Given no prior information of the target field, it has been shown that a regular beam array, i.e. equiangular projections and equispaced laser beams within each projection, leads to uniform distribution of the sampling deficiency across the RoI [32]. Therefore, a parallel beam arrangement is adopted in this work, as shown in Fig. 3, with 32 laser beams arranged in 4 equiangular projection angles, each angle with 8 equispaced parallel beams. The same optical layout is used for both the simulation and the experiment.

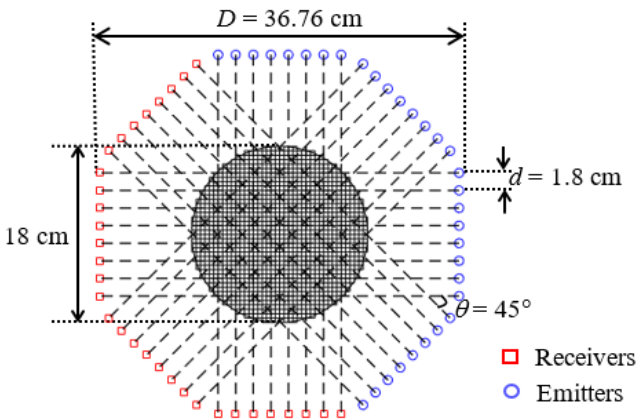

Figure 3: Optical layout of the TDLAS tomographic sensor.

The angular spacing between projections, $\theta$, is 45 degrees. The neighbouring beam spacing, $d$, is 1.8 cm, while the distance $D$ between each emitter and detector is 36.76 cm. The central RoI is defined as a circular sensing area with diameter 18 cm that encloses the highest density of beam crossings, as shown in Fig. 3. The dimensions of each pixel in the RoI are 0.225 cm × 0.225 cm, resulting in 1396 uniformly segmented pixels.

Three phantoms of two-dimensional (2D) distributions of temperature and $H_2O$ concentration are generated with one, two and three inhomogeneities, respectively. As shown in Fig. 4, each inhomogeneity is simulated by a 2D Gaussian profile. In a hydrocarbon flame, the $H_2O$ concentration is generally well-correlated with the temperature. Therefore, the $H_2O$ concentration distribution in each phantom is similar to the temperature distribution. The distributions of temperature and $H_2O$ concentration are mathematically expressed as

$$T(x,y) = 298.15 + \sum_{k=1}^{K} 800 \exp\left[-\frac{(x-x_c^k)^2 + (y-y_c^k)^2}{\sigma^2}\right], \quad (17)$$

$$X(x,y) = \sum_{k=1}^{K} 0.1 \exp\left[-\frac{(x-x_c^k)^2 + (y-y_c^k)^2}{\sigma^2}\right], \quad (18)$$

where $x$ and $y$ denote the horizontal and vertical coordinates of the RoI respectively. $(x_c^k, y_c^k)$ is the central position of the $k$-th Gaussian profile. $\sigma$ is the standard deviation. Table 1 details the parameters of the three phantoms. A high-resolution RoI with 13096 pixels, each with size 0.09 cm × 0.09 cm, is used in the phantoms to calculate the path integrals $A_v$. As a result, the forward problem in Eqn. (3) can be formulated with high accuracy, thus facilitating the analysis of noise performance in Section III.C.

Table 1: Simulation parameters of the three phantoms in Fig. 4.

| | $(x_c^k, y_c^k)$ [cm] | $\sigma$ [cm] |
|---|---|---|
| Phantom 1 | (0, 0) | 0.9 |
| Phantom 2 | (-3.82, -3.82) (3.82, 3.82) | 0.45 |
| Phantom 3 | (-2.7, 4.68) (-2.7, -4.68) (5.4, 0) | 0.45 |

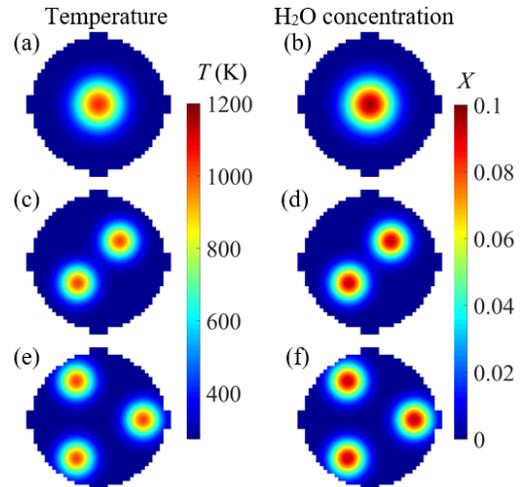

Figure 4: Phantoms of 2D distribution of temperature and $H_2O$ concentration with (a, b) one, (c, d) two, and (e, f) three inhomogeneities, respectively.

B. Metrics for image quality quantification

To examine the performance of the reconstruction algorithms, the tomographic images can be evaluated quantitatively using the following four metrics:

**Image Error (IE):** IE is defined as the relative error between the reconstructed and the true image, indicating the overall quality of the reconstructed image.

$$\text{IE} = \frac{\|T_{\text{rec}} - T_{\text{true}}\|_2}{\|T_{\text{true}}\|_2}, \tag{19}$$

where $T_{\text{rec}}$ and $T_{\text{true}}$ represent the reconstructed and the true temperature distributions, respectively.

**Dislocation (DL):** DL characterises the relative error of the centroid locations between the reconstructed inhomogeneity $(x_r, y_r)$ and that of true inhomogeneity $(x_c, y_c)$. The centroid of the Gaussian-shaped inhomogeneity in the phantoms coincides with its centre.

$$\text{DL} = \frac{\sqrt{(x_r - x_c)^2 + (y_r - y_c)^2}}{r_{\text{RoI}}}, \tag{20}$$

where $r_{\text{RoI}}$ is the radius of the RoI. For phantoms with multiple inhomogeneities, the mean value of DL is given below.

**Centroid Value Error (CVE):** CVE calculates the relative difference of temperature values at the centroids of the reconstructed inhomogeneity $T_{\text{rec}}(x_r, y_r)$ and that of the true inhomogeneity $T_{\text{true}}(x_r, y_r)$.

$$\text{CVE} = \frac{|T_{\text{rec}}(x_r, y_r) - T_{\text{true}}(x_c, y_c)|}{|T_{\text{true}}(x_c, y_c)|}. \tag{21}$$

For phantoms with multiple inhomogeneities, the mean value of CVE is given below.

**Overshoot (OS):** OS is defined as the ratio of the number of pixels assessed as outliers, $N_{\text{OT}}$, to the total number of pixels, $N$.

$$\text{OS} = \frac{N_{\text{OT}}}{N}. \tag{22}$$

In a window containing $3 \times 3$ pixels, the mean temperature, $\overline{T_w}$, and the standard deviation, $\sigma_w$, is calculated. The outliers are defined as the pixels within the window with temperature values deviating from $\overline{T_w}$ by more than $3 \times \sigma_w$. $N_{\text{OT}}$ for a whole image is obtained from accumulating the number of outliers when the window moves across the RoI. Therefore, the extent of spike noise in the tomographic image can be characterised by OS.

C. Determination of the regularisation parameters

As noted in Section II, the regularisation parameters play an important role in both SART and the proposed RETRO. One regularisation parameter, $\lambda$, is involved in SART, while two, $\gamma$ and $\mu$, are used in RETRO. In this subsection, these regularisation parameters are optimally determined using the phantoms in Fig. 4. For other applications with very different phantoms, the regularisation parameters can be determined by following the method detailed below.

In the simulation, the SNR of the LOS-TDLAS measurements was set to 40 dB, which is similar to the noise performance in the experiment. IE, the metric defined above to evaluate overall image quality, was calculated for all three phantoms when the regularisation parameters were varied over wide ranges. For each given value of the regularisation parameter, IE was averaged for 20 repeated simulations. The optimal regularisation parameters were selected as those where the minimum value of the averaged IE was obtained, i.e. when the best image quality was achieved.

For SART, the averaged IE was calculated when $\lambda$ varies from 1 to $10^{-9}$ with 40 steps of logarithmic decrement. As shown in Fig. 5, the dependence of IE on $\lambda$ follows the same trend for each of the three phantoms. IE decreases as $\lambda$ decreases from 0.2. When $\lambda$ equals 0.1, the minimal values of IE are 0.13, 0.19, and 0.28 for phantoms 1, 2, and 3, respectively. IE gradually increases as $\lambda$ further decreases from 0.1. Therefore, the optimal $\lambda$ is selected as 0.1 in SART.

For RETRO, the two regularisation parameters $\gamma$ and $\mu$ were jointly evaluated to achieve the optimal image quality. For each phantom, the averaged IE was calculated for 100 different combinations of $\gamma$ and $\mu$ when each parameter varies from $10^{-1}$ to $10^{-9}$. As shown in Fig. 6, relatively small values of IE can be obtained for all three phantoms within the parameter set $\{(\mu,\gamma): \mu \leq 10^{-2}\gamma, 10^{-5} \leq \gamma \leq 10^{-1}, 10^{-6} \leq \mu \leq 10^{-2}\}$. The optimal $\gamma$ and $\mu$ are 0.01 and $10^{-5}$, respectively, giving the values of IE 0.10, 0.12 and 0.14 for the three phantoms. This optimal combination of $\gamma$ and $\mu$ in RETRO is highlighted in Fig. 6.

D. Results and discussion

With the regularisation parameters determined above, both SART and RETRO were used to reconstruct simulated data from the three temperature images shown in Fig. 4. The simulation was firstly implemented with TDLAS tomographic data at the SNR of 40 dB. As this work focuses on substantial improvement on temperature imaging in TDLAS tomography, we only show the reconstructed temperature images to avoid readers' distraction from the main focus. The improved quality of temperature images will also contribute to a better accuracy of the gas concentration distributions that can be subsequently solved by linear tomographic algorithms [15].

Fig. 7 shows the three reconstructed temperature images. The proposed RETRO algorithm outperforms SART for all three phantoms. Although both algorithms are capable of locating and displaying the inhomogeneities, fewer artefacts are observed in the images reconstructed using RETRO.

To achieve quantitative validation of the proposed new algorithm, the four metrics described in Section II B were calculated, with results presented in Table 2. In terms of overall image quality, the proposed RETRO algorithm improves 4% - 15% on IE compared with SART. In addition, RETRO is much superior in localising inhomogeneities in the reconstructed temperature images since DL values obtained using RETRO, less than 1.2% for all three phantoms, are approximately half of the values obtained using SART. A slightly better performance on CVE can be seen using RETRO, although both algorithms estimate temperature values at the centroids with all CVEs larger than 5%, arguably a poor level of accuracy. This is possibly caused by the smoothness regularisation imposed on both algorithms, leading to robustness in image reconstruction at the cost of accuracy and bias errors in the retrieved temperature values. Finally, RETRO is much better at suppressing spike noise with smaller OS than those obtained using SART. In particular, OS for phantom 3 is 1.06% using the RETRO, which is approximately four times less than that using SART.

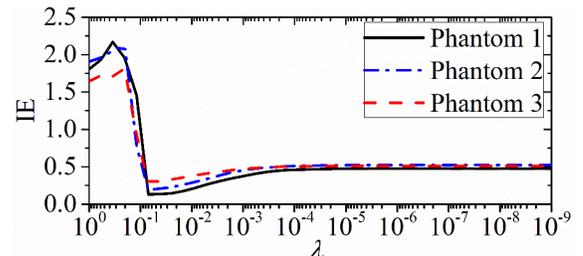

Figure 5: Dependence of IE on $\lambda$ in the SART.

The performances of the two algorithms were evaluated further using simulated TDLAS measurements contaminated with different noise levels. The dependence of all the four metrics at different SNRs are shown in Figure 8. For a given SNR, the value of each metric is averaged for the three phantoms. For the metrics of IE, DL and OS, the results obtained using RETRO are persistently better than those using SART. For CVE, RETRO introduces more significant improvement in relatively low-SNR scenarios, which commonly exist in industrial field applications.

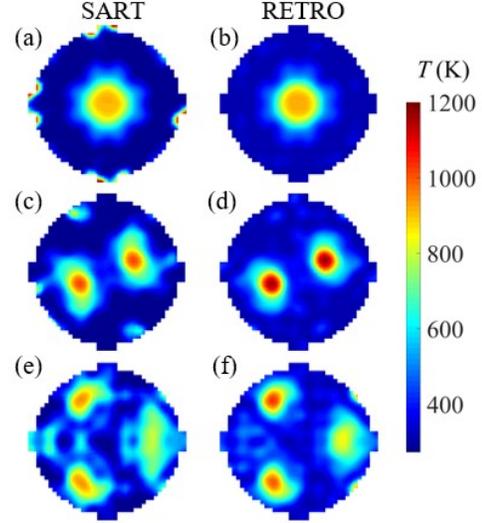

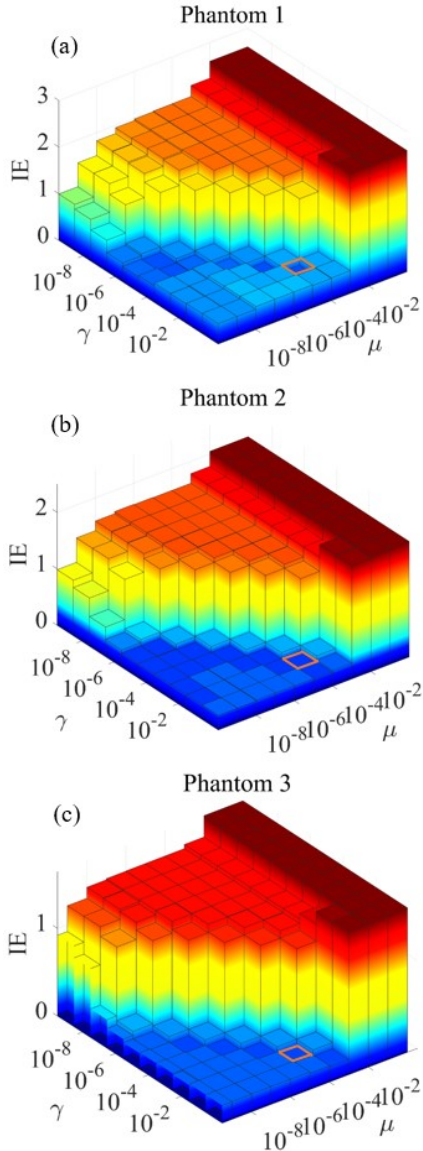

Figure 6: Dependence of IE on $\gamma$ and $\mu$ for (a) phantom 1, (b) phantom 2, and (c) phantom 3 using the proposed RETRO algorithm.

Figure 7: Reconstructed temperature images using SART and the proposed RETRO algorithms with TDLAS tomographic data at the SNR of 40dB for (a, b) phantom 1, (c, d) phantom 2, and (e, f) phantom 3, respectively.

Table 2. Quantitative evaluation of the SART and the proposed RETRO using the four metrics proposed in Section II B.

|  |  | Phantom1 | Phantom2 | Phantom3 |
|---|---|---|---|---|
| IE (%) | SART | 14.7 | 23.6 | 31.8 |
|  | RETRO | 10.1 | 12.3 | 16.9 |
| DL (%) | SART | 1.07 | 1.93 | 1.97 |
|  | RETRO | 0.55 | 1.15 | 1.05 |
| CVE (%) | SART | 8.64 | 7.9 | 15.1 |
|  | RETRO | 8.31 | 5.61 | 12.9 |
| OS (%) | SART | 2.35 | 1.92 | 4.17 |
|  | RETRO | 1.53 | 1.17 | 1.06 |

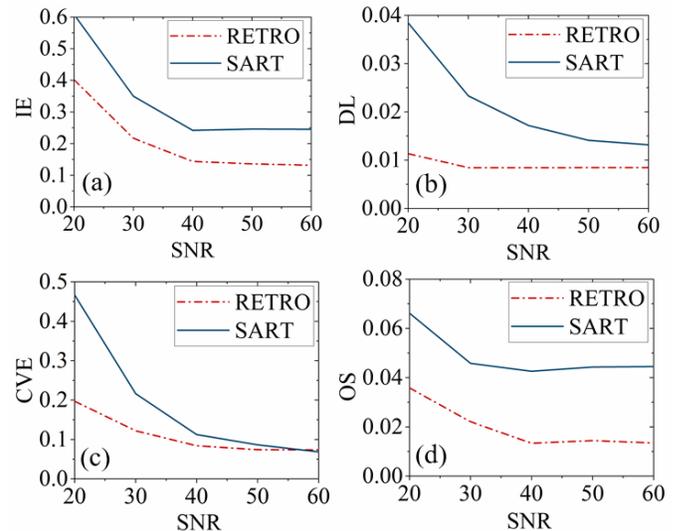

Figure 8: Dependence of the four proposed metrics, (a) IE, (b) DL, (c) CVE, and (d) OS, on different SNRs for both SART and RETRO.

## IV. EXPERIMENTAL VALIDATION

In this section, lab-scale experiments were carried out to further validate the proposed RETRO algorithm. The TDLAS tomographic sensor was built with the same optical layout as depicted in Fig. 4. The lasers at the transitions $v_1$=7185.6 cm$^{-1}$ and $v_2$=7444.36 cm$^{-1}$ were provided by two distributed feedback laser diodes NLK1E5GAAA and NLK1B5EAAA (NTT Electronics), respectively. Current modulation with a scan frequency $f_s$=200 Hz and a modulation frequency $f_m$=40 kHz was imposed on both laser diodes. With the time division multiplexing scheme between the two lasers, a temporal resolution of 10 ms can be achieved in the proof-of-concept experiment. The two laser diodes with pigtailed fibres were combined and split by a 2 × 32 fibre coupler to deliver the 32 laser beams in the tomographic sensor. Each collimated laser beam was detected by an InGaAs photodiode (G12182-010K, Hamamatsu) and then digitised by a data acquisition platform (RedPitaya) at 3.9 Mega Samples/second. Digital lock-in filters were used to extract the Wavelength Modulation Spectroscopy (WMS) first and second harmonic signals, i.e. 1$f$ and 2$f$, from the transmitted signals. Finally, the calibration-free 1$f$ normalised 2$f$ signal, i.e. WMS-2$f$/1$f$ [33, 34], was fitted to calculate each of the 32 path integrals, i.e. the measured data, for each spectral transition.

In the experiment, three temperature distributions were generated to validate the proposed RETRO algorithm. As shown in Fig. 9, there is a single flame located at the centre of the RoI in case 1. In case 2, the same flame is relocated at the lower centre of the RoI. A more complex temperature profile with two flames in the RoI is used in case 3. Estimates of all three real temperature images were reconstructed from the same frame of measurement data using both RETRO and SART.

As shown in Fig. 10, the three temperature images reconstructed using RETRO have fewer artefacts and less spike noise than those using SART. For case 1, the location of the flame reconstructed using RETRO agrees well with the original one shown in Fig. 9 (a), while that obtained from SART is distorted to the right of the centre. The peak temperature values of the inhomogeneity retrieved by RETRO and SART in case 1 are 713 K and 816 K, respectively. For case 2, the relocated flame can be clearly localised using RETRO, the peak value of the retrieved temperature now being 683 K, similar to that in case 1. However, the temperature image in case 2 reconstructed by SART suffers from significant artefacts at the upper centre of the RoI, showing an unreal high-temperature inhomogeneity with maximum temperature value of 1159 K. In addition, the peak temperature value of the real inhomogeneity reconstructed using SART is 574 K, an offset of 242 K compared with that in case 1. For case 3, RETRO can reliably resolve the two flames with their locations being consistent with those in the real temperature distribution. However, the boundaries of the flames retrieved by SART are severely blurred with much lower temperature values.

## V. CONCLUSION

In this paper, we developed a radically new reconstruction algorithm for high-fidelity temperature imaging in TDLAS tomography, called the Relative Entropy Tomographic RecOnstruction (RETRO) algorithm. RETRO enables temperature retrieval from the jointly reconstructed two-line absorbance distributions, thus significantly suppressing the spike noise caused by the ill-posed nature of TDLAS tomography. RETRO was analytically demonstrated to be successful in retaining useful details of the reconstructed temperature image while maintaining good robustness against noise.

The performance of the proposed RETRO algorithm was compared with the SART algorithm, by simulation and by experiment, using two $H_2O$ transitions and a fixed beam arrangement. The regularisation parameters for both algorithms were optimised using noise-contaminated data with the SNR of 40 dB. Four metrics were proposed to evaluate quantitatively the performances of both algorithms. For a wide range of measurement SNR, simulation results indicate that RETRO can reconstruct the temperature images for the three phantoms used with better quality in comparison with SART. In the experiment, both algorithms were used to reconstruct three different temperature distributions. Compared with those obtained using SART, the tomographic temperature images using RETRO achieved much better agreement with the known locations of the original flames and have stronger resistance to spike noise.

In summary, both simulation and experiment indicate that the proposed new RETRO algorithm for two-line temperature tomography is spatially accurate and robust to noisy measurements, and is superior to well-established algorithms for TDLAS tomography. These attributes are invaluable for implementations in harsh environments. In our future work, the RETRO algorithm will be enhanced to incorporate the redundancy of temporal information to investigate the dynamic evolution of the flame.

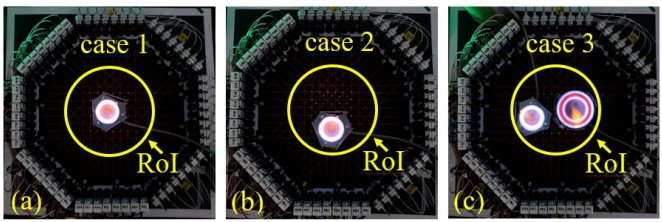

Figure 9: Three temperature distributions generated in the experiment.

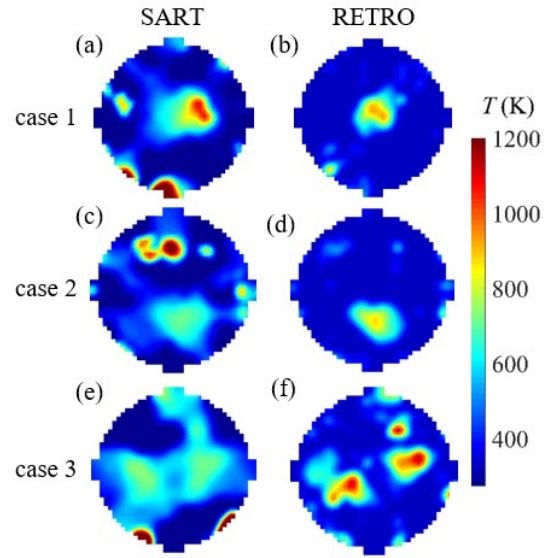

Figure 10: Temperature images reconstructed using both the SART and the proposed RETRO for (a, b) case 1, (c, d) case 2, and (e, f) case 3, respectively.

ACKNOWLEDGEMENT

The authors gratefully acknowledge financial support from the UK Engineering and Physical Sciences Research Council (Platform Grant EP/P001661/1), and from the European Union (H2020 contract JTI-CS2-2017-CFP06-ENG-03-16). The authors would also like to thank for Dr. Nick Polydorides for his useful suggestions.